\documentclass[%
 aip,
 amsmath,amssymb,
 reprint,
]{revtex4-1}

\usepackage{graphicx}
\usepackage{dcolumn}
\usepackage{bm}

\usepackage[utf8]{inputenc}
\usepackage[T1]{fontenc}
\usepackage{mathptmx}
\usepackage{etoolbox}
\usepackage{xcolor}

\makeatletter
\def\@email#1#2{%
 \endgroup
 \patchcmd{\titleblock@produce}
  {\frontmatter@RRAPformat}
  {\frontmatter@RRAPformat{\produce@RRAP{*#1\href{mailto:#2}{#2}}}\frontmatter@RRAPformat}
  {}{}
}%
\makeatother
\begin{document}


\title[Adaptive Exponential Integrate-and-Fire Model with Fractal Extension]{Adaptive Exponential Integrate-and-Fire Model with Fractal Extension}
\author{Diogo L. M. Souza}
\affiliation{ Graduate Program in Science, State University of Ponta Grossa, 84030-900 Ponta Grossa, PR, Brazil}

\author{ Enrique C. Gabrick}
\affiliation{ Graduate Program in Science, State University of Ponta Grossa, 84030-900 Ponta Grossa, PR, Brazil}
 \affiliation{Department of Physics, Humboldt University Berlin, Newtonstra{\ss}e
15, 12489 Berlin, Germany.}
\affiliation{Potsdam Institute for Climate Impact Research, Telegrafenberg A31,
14473 Potsdam, Germany.}
 
\author{Paulo R. Protachevicz}
\affiliation{Institute of Physics, University of São Paulo, 05508-090 São Paulo, SP, Brazil}

\author{ Fernando S. Borges}
 \affiliation{ Department of Physiology and Pharmacology, State University of New York Downstate Health Sciences University, Brooklyn, New York 11203, USA }
 \affiliation{ 
  Center for Mathematics, Computation, and Cognition, Federal University of ABC, 09606-045 São Bernardo do Campo, SP, Brazil\\}
  
\author{ Jos\'e Trobia }
 \affiliation{ Department of Mathematics and Statistics, State University of Ponta Grossa 84030-900, Ponta Grossa, Brazil }
 
 \author{ Kelly C. Iarosz  }
\affiliation{
University Center UNIFATEB, 84266-010, Telêmaco Borba, PR, Brazil }

\author{ Antonio M. Batista }
 \affiliation{ Graduate Program in Science, State University of Ponta Grossa, 84030-900 Ponta Grossa, PR, Brazil}
 \affiliation{ Department of Mathematics and Statistics, State University of Ponta Grossa 84030-900, Ponta Grossa, Brazil }
\affiliation{Institute of Physics, University of São Paulo, 05508-090 São Paulo, SP, Brazil }

\author{
 Iber\^e L. Caldas }\affiliation{Institute of Physics, University of São Paulo, 05508-090 São Paulo, SP, Brazil }

\author{Ervin K. Lenzi }
\affiliation{ Graduate Program in Science, State University of Ponta Grossa, 84030-900 Ponta Grossa, PR, Brazil}
 \affiliation{ 
Departament of Physics, State Univesity of Ponta Grossa, Av. Gen. Carlos Cavalcanti 4748,
Ponta Grossa 84030-900, PR, Brazil}
  \email{protachevicz@gmail.com}

\begin{abstract}
The description of neuronal activity has been of great importance in neuroscience. In this field, mathematical models are useful to describe the electrophysical behaviour of neurons. One successful model used for this purpose is the Adaptive Exponential Integrate-and-Fire (Adex), which is composed of two ordinary differential equations. Usually, this model is considered in his standard formulation, i.e., with integer order derivatives. In this work, we propose and study the fractal extension of Adex model, which in simple terms corresponds to replacing the integer derivative by non-integer. As non-integer operators, we choose the fractal derivatives. We explore the effects of equal and different orders of fractal derivatives in the firing patterns and mean frequency of the neuron described by the Adex model. Previous results suggest that fractal derivatives can provide a more realistic representation due to the fact that the standard operators are generalized. Our findings show that the fractal order influences the inter-spike intervals and changes the mean firing frequency. In addition, the firing patterns depend not only on the neuronal parameters but also on the order of respective fractal operators. As our main conclusion, the fractal order below the unit value increases the influence of the adaptation mechanism in the spike firing patterns.
\end{abstract}

\maketitle

\begin{quotation}
Non-integer derivatives are able to provide a more realistic representation of different dynamical systems activities due to the generalization of the derivative operator. Motivated by this, we study the dynamic behaviour of the adaptive exponential integrate-and-fire model with derivative fractal extension. We observe that the neuron dynamics depend on the fractal order. Decreasing
the fractal order, the adaptation and the coefficient of variation increase, as well as the firing frequency reduces. For some specific values of reset parameters, the fractal order plays a crucial role in the firing pattern. In our simulations, we show that the firing pattern is not only dependent of the reset conditions, as the standard model, but it also has a dependency on the derivative fractal order of the membrane potential and adaptation current.
\end{quotation}

\section{Introduction}

The description of the membrane potential is a fundamental question in neuroscience~\cite{Northrop2001}. Mathematical models have been used to investigate this subject~\cite{ref-neuronalDynamics2014}, such as Hodgkin-Huxley \cite{ref-Hodgkin}, Hindmarsh-Rose \cite{ref-HindmarshRose}, Izhikevich \cite{ref-Izhikevich}, Lapicque integrate-and-fire (IF) \cite{ref-lacpique}, adaptive exponential integrate-and-fire (Adex) \cite{ref-adex2005}, Rulkov map \cite{ref-RulkovMap}, among others \cite{ref-Borges2022, Viana2014, ref-Borges2015, Copelli2006, Sayari2022}. The differences among these models are in the mathematical framework, the biological description \cite{ref-neuronalDynamics2014}, and the computational cost \cite{Sayari2023}. Usually, models with greater biological fidelity require many parameters and high computational costs. It is possible to describe the membrane potential in a really simplified way, however, sometimes with no satisfactory biological meaning. In this work, we consider the Adex model that has a good balance between the biological meaning, computational process cost, and a relatively small set of parameters \cite{Borges2023, ref-firingAdex, ref-SincronizationAdex}. 

Adex model was proposed in 2005 by Brette and Gerstner as an improvement of the leak conductance neuronal model \cite{ref-adex2005}. Such improvement consists of the inclusion of a spike mechanism term in the potential variable and an additional variable to describe the adaptation mechanism. Besides the simplicity of this model and its low computational cost, it presents a great biophysical accuracy \cite{ref-neuronalDynamics2014} and fits neuronal dynamics \cite{ref-Hertag2012}. In the Adex, the membrane potential depends on the adaptation current that describes the slow activation and deactivation of some potassium ionic channels in the neuron \cite{ref-Brette2015}. A spike threshold mechanism, represented by an exponential term, describes the fast arising when an action potential is generated \cite{ref-neuronalDynamics2014, ref-adex2005, ref-firingAdex}. However, the exponential term introduces a discontinuity in the model which is solved considering the reset condition when the membrane potential overpasses a certain threshold. Depending on the reset condition in the potential and adaptation current, it is possible to reproduce different firing patterns \cite{ref-firingAdex}. Moreover, for some sets of these parameters, chaotic solutions are found \cite{ref-BifurcationAdex}. 

The standard description of the Adex model is based on ordinary differential equations (ODE), where the differential operators have integer order. This description has good accuracy in describing real patterns \cite{ref-firingAdex} and exhibits rich dynamic solutions of the neuronal activities \cite{Protachevicz2019}. Nonetheless, recent developments have been showing that extensions of integer operators to non-integer can increase the accuracy of the models to fit real data \cite{Hernandez2021} and modify the dynamical properties \cite{Jafari2021}. The most famous non-integer extensions are the fractional operators. Fractional calculus has been used in many fields, such as quantum mechanics \cite{Gabrick2023, Lenzi2023}, Hamiltonian systems \cite{Tarasov2005}, photothermal \cite{Somer2023}, epidemiology \cite{Pinto2019}, and others \cite{Tarasov2020, Arora2022, Giusti2020, ref-PorousMedia, ref-AnomalousDiffusion, ref-HeatConduction, ref-DarkEnergy}. In the neuroscience models, fractional extensions of Hindmarsh–Rose \cite{Jun2013, Yu2020}, Hodgkin–Huxley \cite{Nagy2014, Weinberg2015}, Rulkov \cite{Lu2022, Ma2023}, and leaky integrate-and-fire \cite{Teka2017, Teka2014} have been studied. However, the literature about fractional extensions of the Adex model is scarce. We address this lack of literature to the discontinuity of integrate-and-fire models and the difficulty to take this particularity in account in fractional derivatives.  Due to fractional calculus properties, it is very hard to work with non-smooth systems. As an alternative of non-integer operators  that we can employ are the fractal ones. The fractal derivative was proposed as local operators~\cite{Chen2006}, which are directly connected with the fractal dimension~\cite{Chen2017}. In this way, one form to understand the non-integer effects in Adex model is by means of fractal derivatives.

Fractal calculus has been considered to describe many phenomena when the standard calculus fails \cite{ref-fractalcalculus}, which is based on the fractal space-time concept \cite{ref-DarkEnergy, ChenBook, Microphysics}. In the context of porous media, where the space is discontinuous, the fractal framework has presented a great description of phenomena \cite{ref-SorbingSystems1, ref-SorbingSystems2, ref-SorbingSystems3}. In theoretical physics, fractal calculus is well explored in dark energy topics \cite{ref-DarkEnergy}. El Naschie considered the fractal space-time and pointed out that dark energy is around $95.5\%$ of the total energy-mass of the Universe \cite{ref-Darkenergy1, ref-Darkenergy2}.  In the context of diffusion process, fractal calculus is used to study anomalous relaxation process \cite{Chen2006, ref-AnomalousDiffusion}. In biological models, the fractal derivatives describe very well the heat conduction in the polar bear hairs \cite{ref-HeatConduction}. In addition, when fractal derivatives are considered in the SIS epidemiological model, the description of real data increases. For example, considering Brazilian data from syphilis, it is possible to obtain a correlation coefficient equal to 0.998 with fractal operators \cite{Gabrick2023b} meanwhile the integer derivative operator produces a correlation coefficient equal to 0.990. Other applications are found in Ref. \cite{Atangana2019}, where the authors showed that for some non-integer order differential operators new system properties emerge.

In this work, we study the behaviour of the Adex model when it is described by fractal order differential equations. We investigate the effect of fractal order $(\alpha)$ with equal values in both neuronal variables ($V$ and $w$) as well as in independent order in each variable. For small values of $\alpha$, we show that the inter-spike intervals (${\rm ISI}$) increase. The change in $\overline{\rm ISI}$ is proportional to an exponential function of $\alpha$. Furthermore, we show that the firing pattern changes when potential membrane and the adaptation current are of two different fractal orders.

\section{Integer Adex Model}
The adaptive exponential integrate-and-fire model \cite{ref-adex2005} is described by the following equations
\begin{eqnarray}
\label{eq1}
C\frac{dV}{dt} &=& -g_{\rm L}(V-E_{\rm R})+g_{\rm L}\Delta_{\rm T} \exp \left(\frac{V-V_{\rm T}}{\Delta_{\rm T}}\right)-w+I, \\
\label{eq2}
\tau_w \frac{dw}{dt} &=& a(V-E_{\rm R})-w,
\end{eqnarray}
where $C$ is the membrane capacitance, $V$ is the membrane potential, $t$ is the time, $g_{\rm L}$ is the leak conductance, $E_{\rm R}$ is the rest potential, $\Delta_{\rm T}$ is the slope factor, $V_{\rm T}$ is the threshold potential, $w$ is the adaptation current, $I$ is the injected current, $\tau_w$ is the time constant, and $a$ is the level of sub-threshold adaptation. When $V$ reaches a maximum value ($V_{\rm max}$), the following reset conditions are applied
\begin{eqnarray}
\label{eq3}
V \rightarrow V_{\rm r},\\
\label{eq4}
w \rightarrow w_{\rm r} = w+b.
\end{eqnarray}
For this model, we employ numerical solutions by the Runge-Kutta 4th order method. In our simulations, we use $C=200$ pF, $g_{\rm L}=12$ nS, $E_{\rm R}=-70$ mV, $\Delta_{\rm T} = 2$ mV, $V_{\rm T} = -50$ mV, $I = 512$ nA, $a=2$ nS, $\tau_w=300$ ms, and  $V_{\rm max}=-40$ mV \cite{ref-firingAdex}. The initial conditions are given by $V(0)=E_{\rm R}$ and $w(0)=0$, which correspond to a neuron initially absent of external current in a rest state.

\begin{figure}
\centering
\includegraphics[scale=0.23]{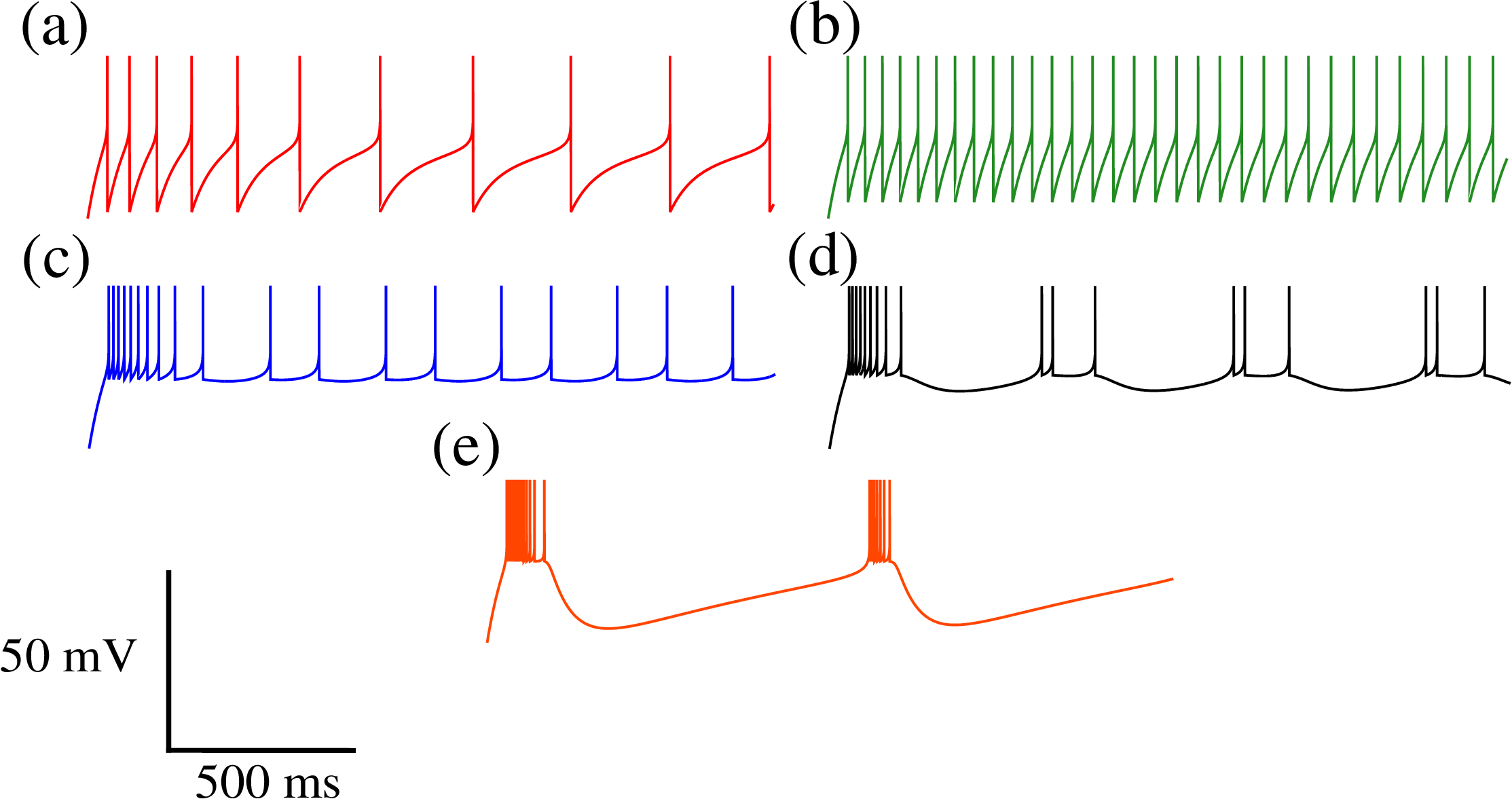}
\caption{Firing patterns generated by the Adex model, using reset parameters on Table \ref{Tab-1}. The panels (a), (b), (c), (d), and (e) display adaptation spiking, tonic spiking, initial bursting, irregular bursting, and regular bursting, respectively.}
\label{Fig1}
\end{figure}

Considering different combinations of $V_{\rm r}$ and $b$, it is possible to distinguish firing patterns \cite{ref-SincronizationAdex}, as displayed in Table \ref{Tab-1}. We identify five firing patterns, adaptation (Figure~\ref{Fig1}(a)), tonic spiking (Figure~\ref{Fig1}(b)), initial bursting (Figure~\ref{Fig1}(c)), irregular bursting (Figure~\ref{Fig1}(d)), and regular bursting (Figure~\ref{Fig1}(e)). In the adaptation dynamics, the inter-spike intervals (ISI) increase throughout the time during the application of a constant current ($I$) due to the adaptive current mechanism. This behaviour is not observed in tonic spiking, where ISI is constant. On the other hand, for initial bursting, the first ISI starts short and then increases over time. For regular bursting, the interval between each bursting train after the transient is constant. Whereas, in irregular bursting the interval between bursting trains is not constant.

\begin{table}[htb]
\caption{Distinguish firing patterns depending on the reset parameters.}
\label{Tab-1}
\centering
 \addtolength{\tabcolsep}{6pt}
  \scalebox{1.1}{
\begin{tabular}{l| c c c}\hline
 \small Firing pattern (Region)  &   \small $V_{\rm r}$ (mV)   & \small  $b$ (pA)  & \small  Fig. \\ \hline
 \small  Adaptation (I)   &  \small  -68.0   &   \small 60   &   \small \ref{Fig1} (a)\\
 \small  Tonic Spiking (II) &  \small  -65.0   &  \small  5  &   \small \ref{Fig1} (b) \\
 \small  Initial Bursting (III)  &  \small -48.8   &  \small  35   &   \small \ref{Fig1} (c) \\
 \small  Irregular Bursting (IV)  &   \small -47.4   &  \small  41   &   \small \ref{Fig1} (d)\\
 \small  Regular Bursting (V)  & \small -45.0   &  \small 40   &  \small \ref{Fig1} (e)\\
  \hline
\end{tabular}}
\end{table}

The ISI is used to distinguish the patterns of spiking and bursting and it is defined by ${\rm ISI}= t_{m+1}-t_{m}$, where $t_m$ is the $m$-th firing of each neuron. In this way, we consider the coefficient of variation (CV) of ISI, CV is given by  
\begin{equation}
\label{eq5}
{\rm CV} = \frac{\sigma_{\rm ISI}}{\overline{\rm ISI}},
\end{equation}
where $\sigma_{\rm ISI}$ is the standard deviation of the $\rm ISI$ normalised by the mean inter-spike interval $\overline{\rm ISI}$ \cite{ref-gabbianiCV}. The bursting pattern produces ${\rm CV}\geq 0.5$ and the spiking pattern produces $\rm CV < 0.5$. In the CV measure, we discard the first four ISIs that are related to transient time. To measure the mean firing frequency, we consider the definition $\overline{F}=\overline{\rm ISI}^{-1}$. 

The adaptive index is employed to distinguish adaptive and tonic spiking. The adaptive index is defined as
\begin{equation}
\label{eq-AdaptiveIndex}
{\rm A} = \frac{1}{q-d-1} \sum_{m=q}^{d} \frac{{\rm ISI}_m - {\rm ISI}_{m-1}}{{\rm ISI}_m + {\rm ISI}_{m-1}},
\end{equation}
where ${\rm ISI}_m$ is the inter-spike interval between the $m$-th and ($m$+1)-th firings. Due to the numerical transient, we discard the first four  ${\rm ISI}$ \cite{ref-firingAdex} by considering $q$=4, and $d$ is the last ${\rm ISI}$. To identify the adaptive or tonic firing pattern, it is necessary to define a critical value of ${\rm A}$, which we define as ${\rm A}_{\rm c} = 0.01$ \cite{ref-firingAdex}. ${\rm A}>{\rm A}_{\rm c}$ and $-{\rm A}_{\rm c} \leq {\rm A} \leq {\rm A}_{\rm c}$ characterise the adaptive and tonic spiking, respectively.

The characterisation of initial, regular, and irregular burstings is done by the analysis of the phase-space ($w\times V$) using the nullcline ($dV/dt=0$) \cite{ref-firingAdex}. Reset condition in the region $dV/dt>0$ is associated with spiking firing pattern while in the region $dV/dt<0$ allows the emergence of bursting firing pattern \cite{Fardet2018}. We count the number of times that $dV/dt>0$ when the reset conditions are applied. If the amount of reset in the region $dV/dt<0$ is the same after resets in the region $dV/dt>0$, the firing pattern is regular bursting, otherwise, we have irregular bursting. If right after the application of the constant current, it is only counted $dV/dt>0$ and after transient time interval we only count $dV/dt<0$, there is an initial bursting. 

\begin{figure}[htb!]
\centering
\includegraphics[scale=0.35]{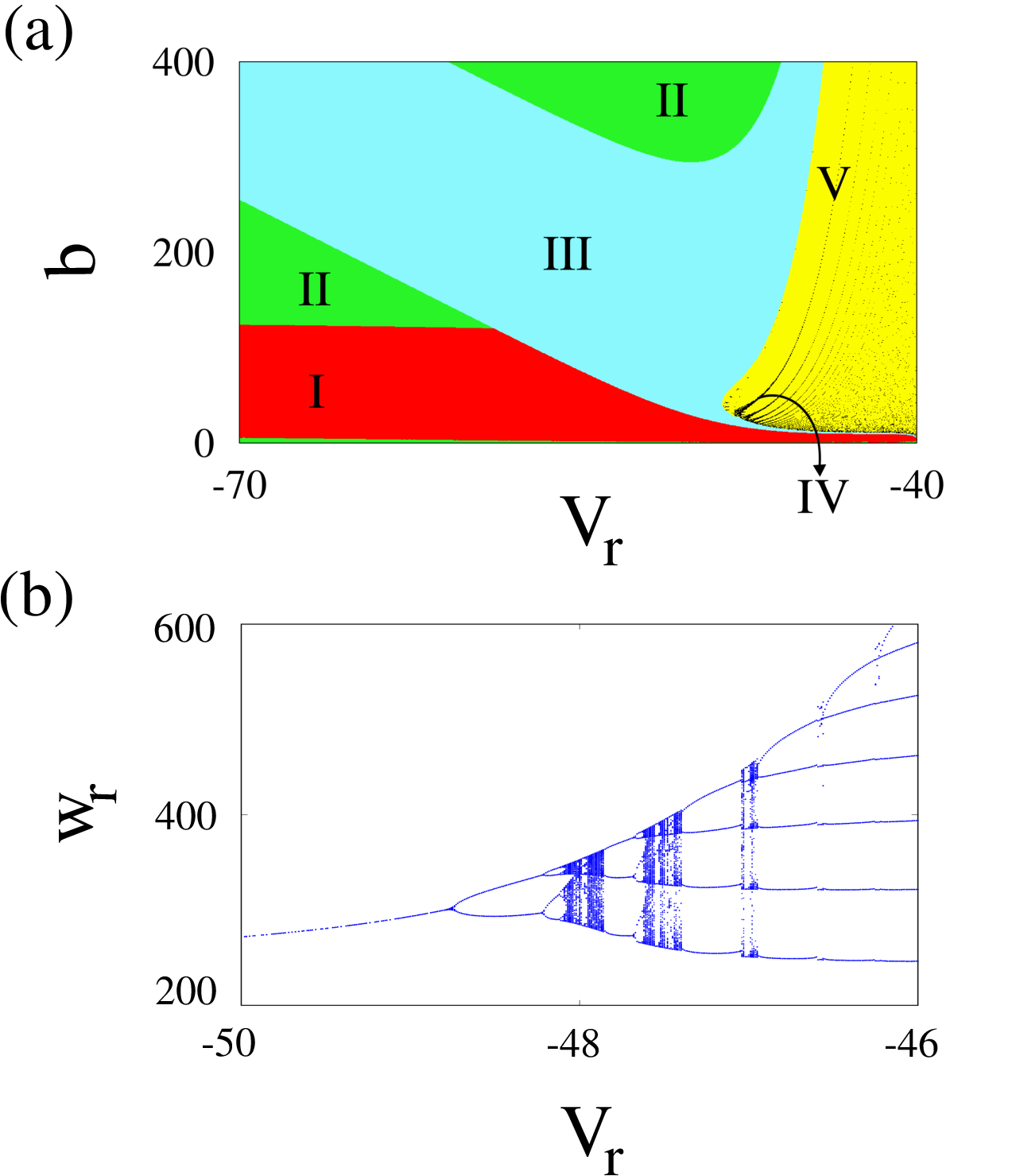}
\caption{Reset parameter space and bifurcation diagram of the Adex model. (a) Parameter space of the firing patterns where each region represents one firing pattern. Regions I, II, III, IV, and V exhibit adaptation spiking, tonic spiking, initial bursting, irregular bursting, and regular bursting firing patterns, respectively. (b) Bifurcation diagram of the Adex model. In the panel (b), we plot $w_{\rm r}$ as a function of $V_{\rm r}$ for $b=80$ pA.}
\label{Fig2}
\end{figure}

Figure \ref{Fig2}(a) displays the reset parameter space $b \times V_{\rm r}$, where each pattern region is identified by different colours. The red colour (Region I) shows the parameters combinations that generate adaptation spiking, the green colour (Region II) exhibits the tonic spiking, the blue colour (Region III) denotes the initial bursting, the black colour (Region IV) exhibits the irregular bursting, and the yellow colour (Region V) corresponds to the parameters that generate the regular bursting. Figure \ref{Fig2}(b) displays the $w_{\rm r}$ bifurcation diagram for different values of $V_{\rm r}$. As $V_{\rm r}$ increases there is a period-doubling bifurcation, with some periodic and chaotic regions.

\section{Fractal Adex Model}
In this section, we present the extension of the adaptive exponential integrate-and-fire model by means of the fractal operator. We study the behaviour of the Adex model when the same fractal order is applied to $V$ and $w$. Moreover, we consider two different fractal orders, $\alpha$ for the fractal order of the membrane potential and $\beta$ for the fractal order of the adaptation current. 

\subsection{Fractal Adex model with equal order}
In this work, we consider the following definition of the fractal operator
\begin{equation}
\frac{df}{dt^\alpha} = \lim_{t_1 \rightarrow t} \frac{f(t_1) - f(t)}{(t_1 - t_0)^\alpha - (t - t_0)^\alpha} = \frac{1}{\alpha (t - t_0)^{\alpha - 1}}\frac{df}{dt},
\end{equation}
as $t_0 = 0$,
\begin{equation}
\label{eq6}
\frac{df}{dt^{\alpha}} = \frac{1}{\alpha}t^{1-\alpha} \frac{df}{dt},
\end{equation}
where $\alpha>0$ is the fractal order. This definition is known as Hausdorff derivative \cite{ref-fractalcalculus}. Considering the model described by Eqs. (\ref{eq1}) and (\ref{eq2}), our proposed extension becomes
\begin{eqnarray}
\label{eq7}
C\frac{dV}{dt^{\alpha}} &=& -g_{\rm L}(V-E_{\rm R})+g_{\rm L}\Delta_{\rm T} \exp \left(\frac{V-V_{\rm T}}{\Delta_{\rm T}}\right)-w+I, \\
\label{eq8}
\tau_w \frac{dw}{dt^{\alpha}} &=& a(V-E_{\rm R})-w.
\end{eqnarray} 
Using the fractal derivative definition Eq. (\ref{eq6}), we rewrite the fractal Adex model as
\footnotesize
\begin{eqnarray}
\label{eq9}
\hspace{-0.1cm} C\frac{dV}{dt}&=& \alpha t^{\alpha-1}\left[-g_{\rm L}(V-E_{\rm R})+g_{\rm L}\Delta_{\rm T} \exp \left(\frac{V-V_{\rm T}}{\Delta_{\rm T}}\right)-w+I\right],\\
\label{eq10}
\tau_w \frac{dw}{dt}&=&\alpha t^{\alpha-1}\left[a(V-E_{\rm R})-w\right],
\end{eqnarray}
\normalsize
where the constants are rewritten in order to preserve the system units. 

\begin{figure}[htb!]
\centering
\includegraphics[scale=0.27]{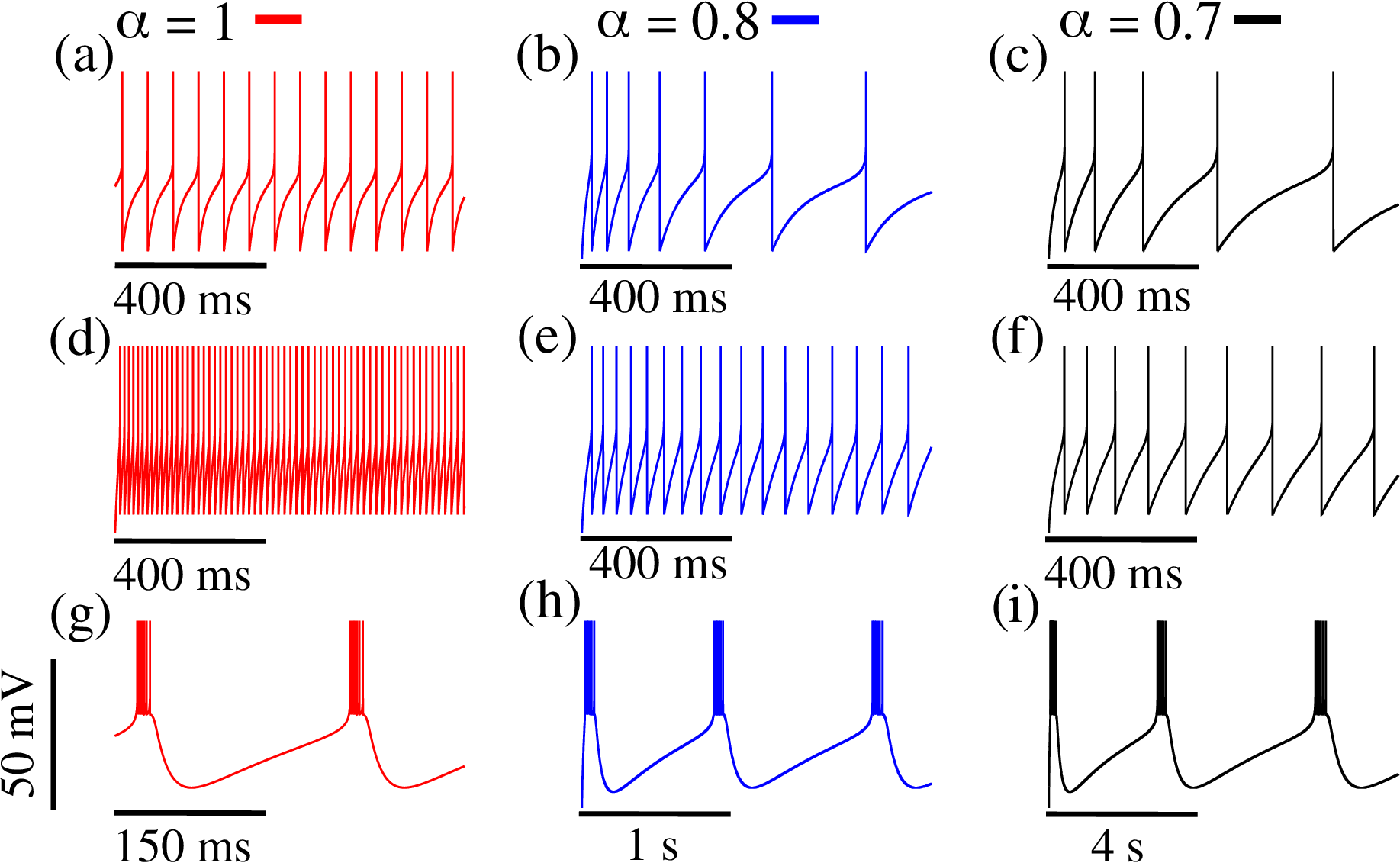}
\caption{Time evolution of the membrane potential of adaptive (a-c), tonic (d-f) and regular burst (g-i) pattern for three different values of $\alpha$. We consider $\alpha = 1$ in the panels (a), (d) and (g), and $\alpha = 0.8$ in the panels (b), (e) and (h) abd $\alpha = 0.7$ in panels (c), (f) and (i). The reset parameters we considered are in table \ref{Tab-1}.}
\label{Fig3}
\end{figure}

We observe that the reduction of the fractal order can increase the effect of the adaptation for spiking patterns. Figures \ref{Fig3}(a-c) show the potential membrane ($V$) as a function of the time ($t$) for $V_{\rm r}=-68$ mV and $b=60$ pA. Decreasing the value of $\alpha$, the value of $\rm ISI$ between the firings increases. Reducing $\alpha$ to even smaller values, the neuron ceases its activity. For $\alpha>1.0$, the intervals are shorter. However, the neuron hyperpolarises to unrealistic values \cite{ref-BrownAndAdams}. To obtain biological solutions, we consider $0.7 \leq \alpha \leq 1.0$. 

For the burst firing pattern, Figure \ref{Fig3} (g-i), the reduction of the fractal order increases the inter-burst intervals. The fractal Adex model for a regular bursting pattern is shown in Figure \ref{Fig3} (g-i). The panels (g) is computed for $\alpha=1$, (h) is computed for $\alpha=0.8$ and (i) is computed for $\alpha=0.7$. The firing pattern does not change by reducing $\alpha$, although the interval between each burst-train changes. The intervals between each burst-train increase considerably.

Most of the firing patterns do not change with $\alpha$ reduction, however, the tonic spiking does change. To show that, we consider the tonic spike firing pattern and reduce the fractal order. Figure \ref{Fig3} (d-f) depict the temporal evolution of the membrane potential. In the panel (d), $\alpha=1$ is associated with the tonic spike of the standard Adex model. The inter-spike intervals are quite similar over time.
In the panels (d-f), the reduction of fractal order to $\alpha=0.7$ generates an adaptation of the spikes over time. Reducing even more the fractal order, generates an increase in the adaptability of the firing patterns, which in this case is characterised by the adaptive spiking. In this way, the tonic firing pattern goes to adaptive spiking, where the fractal order less than the unity acts as an adaptation mechanism in the neuron, increasing the influence of adaptation current in the model.

\begin{figure}[htb!]
\centering
\includegraphics[scale =0.45]{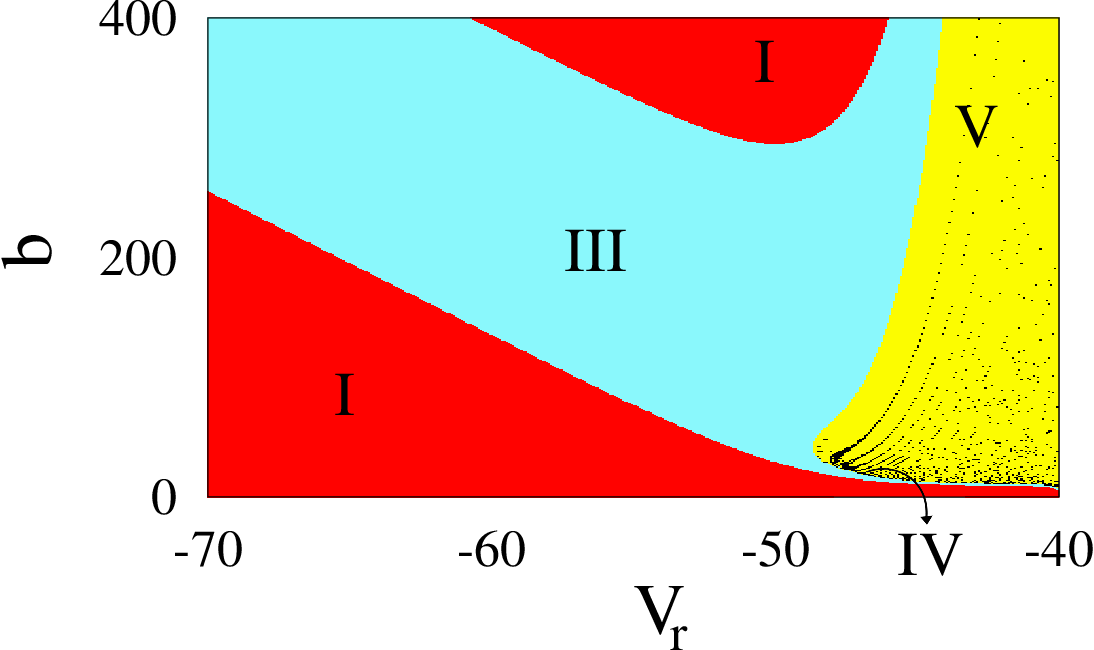}
\caption{Reset parameter space for $\alpha = 0.7$. Region I corresponds to the adaptive spiking, Region III is related to the initial bursting, Region IV is associated with the irregular bursting, and Region V corresponds to the regular bursting.}
\label{Fig4}
\end{figure}

Figure \ref{Fig4} exhibits the reset space parameter for $\alpha =0.7$. The reset parameter space of the standard Adex model (Figure \ref{Fig2}(a)) and reset parameter space of $\alpha=0.7$ (Figure \ref{Fig4}) are very similar in terms of the firing patterns. However, there are not tonic spiking regions in Figure \ref{Fig4}. The tonic spiking becomes adaptive as $\alpha$ reduces. For $0.7 \leq \alpha<1$, the reset parameter space does not change in Figure \ref{Fig4}. For $\alpha=1$,  the reset parameter space is the same as the standard Adex model, displayed in Figure \ref{Fig2}. The sudden change from Figure \ref{Fig2} to Figure \ref{Fig4} is due to the effect of $\alpha$ on the interspike interval, which has a major effect on the tonic spiking region.

\begin{figure}[htb!]
\centering
\includegraphics[scale=0.45]{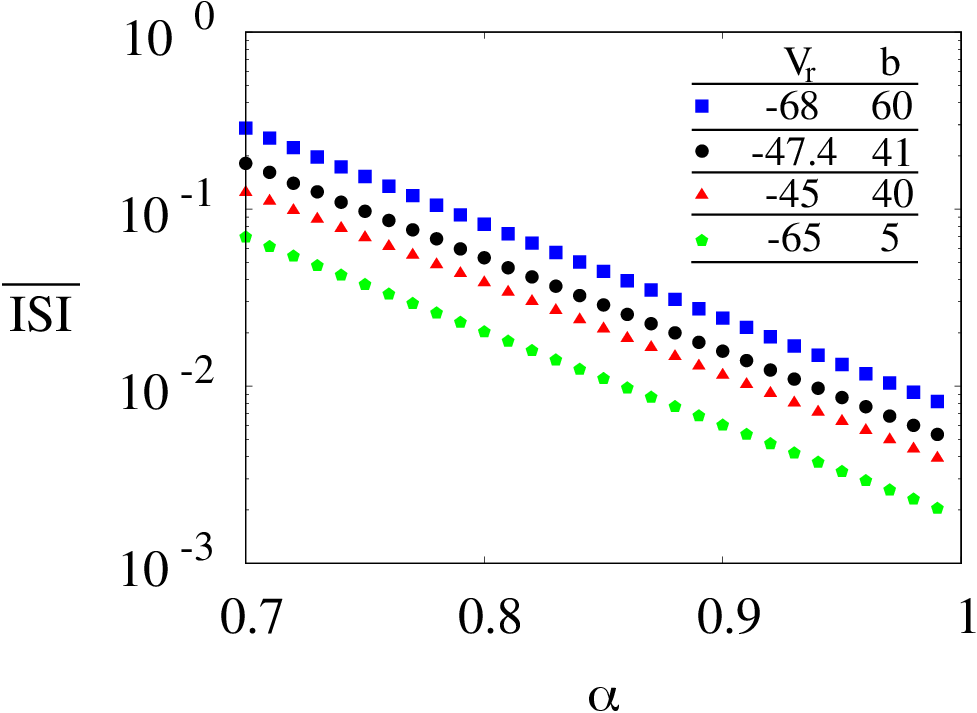}
\caption{$\overline{\rm ISI}$ versus $\alpha$ for different firing patterns. The $\overline{\rm ISI}$ values decrease exponentially with the reduction of $\alpha$. The reset values are shown in Table \ref{Tab-1} and the slopes in Table \ref{Tab-2}.}
\label{Fig5}
\end{figure}

Although the firing pattern does not change for all patterns,  the inter-spike intervals change and depend on the value of $\alpha$, as shown in Figure \ref{Fig3}. Figure \ref{Fig5} displays $\overline{\rm ISI}$ in log scale as a function of $\alpha$ for the firing patterns, distinguished by the colour and point type. The blue squares are for $V_{\rm r} = -68$ mV and $b=60$ pA, the black circles are for $V_{\rm r}=-47.4$ mV and $b=41$ pA, the red triangles are for $V_{\rm r}=-45$ mV and $b=40$ pA, and the green stars are for $V_{\rm r} = -65$ mV and $b=5$ pA. 

Independent of the reset parameters, $\overline{\rm ISI}$ changes in a similar way. Figure \ref{Fig5}
shows that such variation follows an exponential function, $\propto \exp(s\alpha)$, where $s$ is the slope and is displayed 
in Table \ref{Tab-2}. The correlation coefficient indicates that the exponential function describes how $\overline{\rm ISI}$ 
changes as $\alpha$ decreases. The four slopes of lines are very similar, indicating that $\overline{\rm ISI}$ variation does not strongly depend on the reset parameters considered.

\begin{table}[htb]
\caption{Slopes of the reset parameters considered in Figure \ref{Fig5}. \label{Tab-2}}
\centering
\addtolength{\tabcolsep}{6pt}
\scalebox{1.1}{
\begin{tabular}{c|c|c}\hline
{\small  Reset parameter ($V_{\rm r}, b$)}   &   $s$ & {\small Correlation Coefficient}\\ \hline
\small (-68, 60)   &   \small -12.23  & \small 0.99997 \\ \
\small (-47.4, 41) &   \small -12.16  & \small 0.99998\\ \
\small  (-45, 40)   &  \small -11.95  & \small 0.99998\\ 
\small  (-65, 5)   &   \small -12.14 & \small 0.99998\\ \hline
\end{tabular}}
\end{table}

\subsection{Fractal Adex model with two fractal orders} 

Now, we study the model dynamics when the fractal order is considered independently in $V$ and $w$, as well as $\alpha$ and $\beta$. Both fractal orders are in the same range, 
$0.7 \leq \alpha \leq 1$ and $0.7 \leq \beta \leq 1$. The model is given by Eqs. (13) and (14).
\footnotesize
\begin{eqnarray}
\label{eq11}
C\frac{dV}{dt}&=&\alpha t^{\alpha-1}\left[-g_{\rm L}(V-V_{\rm R})+g_{\rm L}\Delta_{\rm T} \exp \left(\frac{V-V_{\rm T}}{\Delta_{\rm T}}\right)-w+I\right], \\
\label{eq12}
\tau_w \frac{dw}{dt}&=&\beta t^{\beta-1}\left[a(V-V_R)-w\right].
\end{eqnarray}
\normalsize

\begin{figure}[htb!]
\centering
\includegraphics[scale = 0.5]{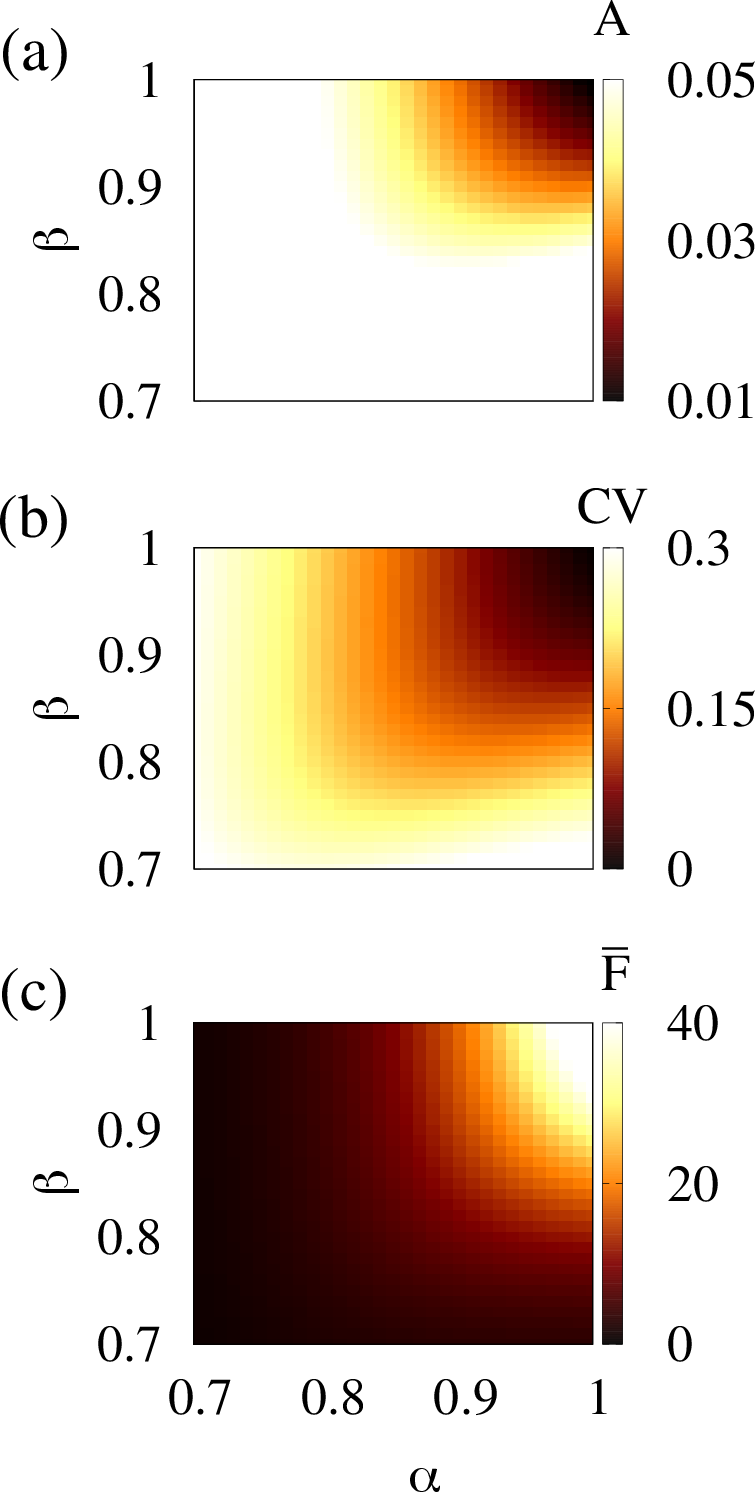}
\caption{Parameter space $\beta\times\alpha$. The colour bars correspond to (a) adaptive index, (b) CV, and (c) mean firing rate. We consider $V_{\rm r}=-65$ and $b=5$, tonic spiking for the standard Adex model.}
\label{Fig6}
\end{figure}

Figures \ref{Fig6} (a-c) display A, CV, and $\overline{F}$, respectively, for tonic spiking reset conditions, considering the parameters shown in Table \ref{Tab-1}. In panel (a), we observe that A increases as $\alpha$ and $\beta$ reduce. A increases at a faster rate with the reduction of $\beta$. The adaptive index reaches greater values than the critical value ($\rm A_c =  0.01$), then characterizing the adaptive spiking firing pattern. To increase spiking adaptability, it is necessary to reduce at least one fractal order. Based on these results, we verify that the fractal order has a large influence on the neuron adapting mechanism. In panel (b), we observe that CV increases with the reduction of the ($\alpha, \beta$), it happens due to the impact of fractal order in the $\overline{\rm ISI}$. However, its increase does not surpass the CV threshold which characterizes bursting (CV$\geq$0.5). Panel (c) shows the mean firing rate parameter space. The reduction of the mean firing rate was expected due to the increase in the $\overline{\rm ISI}$ occasioned by the fractal order.

\begin{figure}[htb!]
\centering
\includegraphics[scale=0.36]{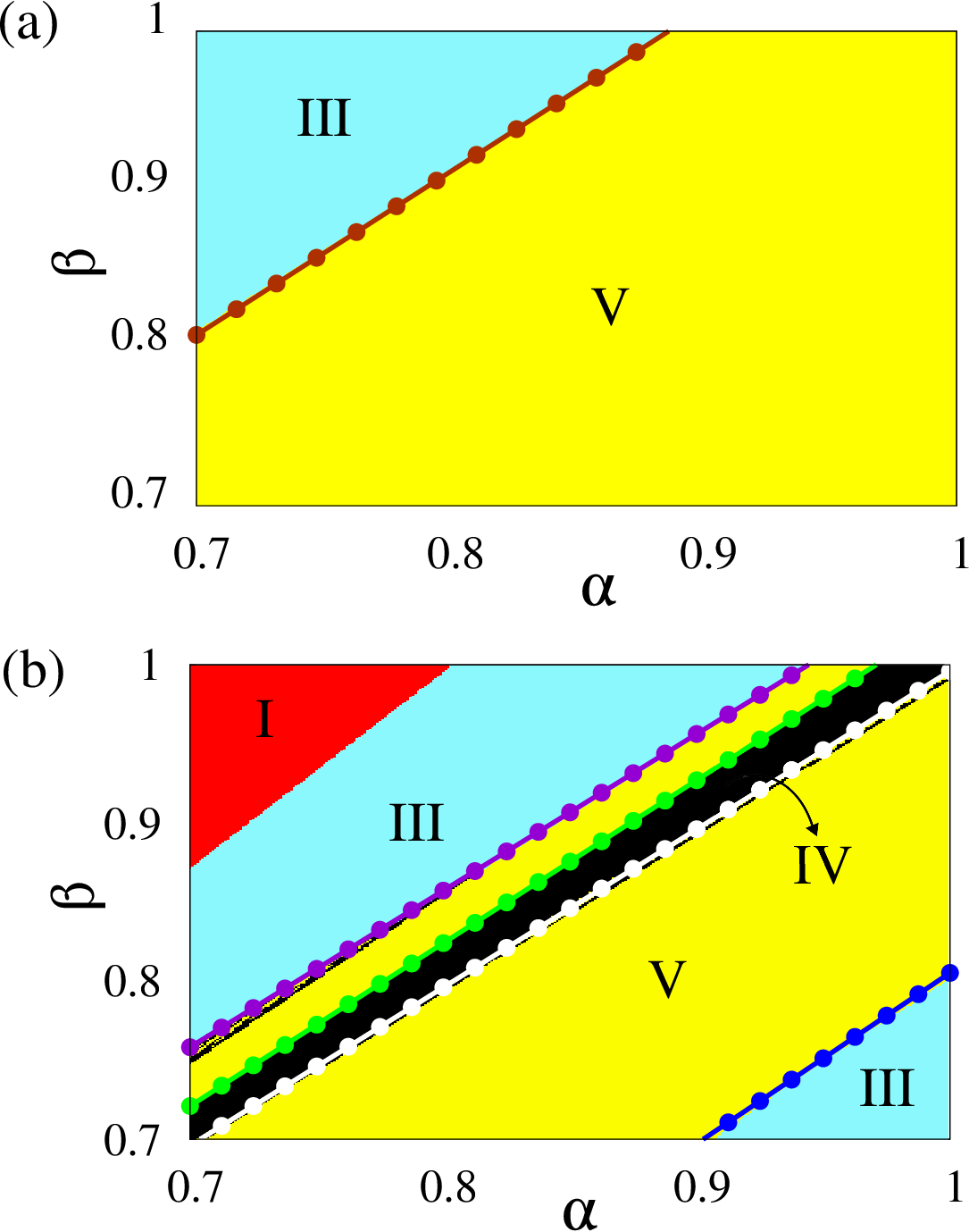}
\caption{Parameter space $\beta\times\alpha$ for (a) $V_{\rm r}=-45$ mV and $b=40$ pA and (b) $V_{\rm r}=-47.4$ mV and $b=41$ pA  according to Table \ref{Tab-1}. Region I shows the adaptive spiking pattern, Region III corresponds to the initial bursting, Region IV exhibits the irregular bursting, and Region V shows the regular bursting. The dotted lines are the boundaries of bursting. The brown dotted line in (a) is the regular bursting boundary. The dotted line in the panel (b) is the boundary of the irregular and regular bursting firing patterns. The purple and blue lines are the boundaries of regular bursting, superior ($g(\alpha)$) and inferior ($q(\alpha)$), respectively. The green and white dotted lines are irregular bursting boundaries, superior ($s(\alpha)$) and inferior ($k(\alpha)$), respectively.
}
\label{Fig7}
\end{figure}

The firing pattern depends on the fractal orders ($\alpha, \beta$) considered as shown in Figures \ref{Fig7} (a) and (b). In panel (a), it is considered $V_{\rm r}=-45$ mV and $b=40$ pA (which characterize regular bursting in the standard model). A transition from regular bursting (Region V) to initial bursting (Region III) is observable. Regular bursting activity is delimited by the brown line given by
\begin{equation}
\label{Eq13}
f(\alpha) = 1.03\alpha+0.088.
\end{equation} 

It is worth mentioning that Eq. (\ref{Eq13}) delimits the transition between regular to initial bursting in Figure \ref{Fig7} (a). This result shows that distinct fractal order, $\alpha$ and $\beta$, change the firing patterns. Therefore, in the extended model, the firing patterns depend not only on the reset parameter but also on fractal orders. For adaptation reset parameters in Figure \ref{Fig2} (a), the space parameter of $\beta \times \alpha$ does not change from the adaptive spike. However, the $\overline{\rm ISI}$ increases. Considering tonic spiking reset parameters in Figure \ref{Fig2} (a), the $\beta \times \alpha$ parameter space shows the transition of tonic to adaptive spiking. Meanwhile, regular bursting reset parameters of Figure \ref{Fig2} (a) produce a transition between firing patterns in the $\alpha \times \beta$ parameter space.

Considering $V_{\rm r}=-47.4$ mV and $b=41$ pA (which characterize irregular bursting in the standard Adex model), the results show four distinct patterns for different pairs of $(\alpha, \beta)$, as shown in Figure \ref{Fig7} (b). Region I (red colour) displays the adaptive spiking, Region III (blue colour) corresponds to the initial bursting, Region IV (black colour) is related to the irregular bursting, and Region V (yellow colour) shows the regular bursting. Different combinations of ($\alpha, \beta$) transform the irregular firing pattern into regular bursting. The firing pattern stays as irregular bursting for pairs of $\alpha$ and $\beta$ within the band defined by the functions $s(\alpha)$ and $k(\alpha)$, green and white line, respectively,
\begin{eqnarray}
k (\alpha) &=& \alpha-0.004,\\
s (\alpha) &=& 1.03\alpha.
\end{eqnarray}
For $k (\alpha)\leq \beta \leq s(\alpha)$, the firing pattern is the same as the case $\alpha=\beta=1$. There are transitions between regular and initial bursting. Initial bursting is easier achieved with the $\alpha$ reduction. Considering $\beta = 1 $, initial bursting is achieved for $\alpha =0.94$. Meanwhile, if we consider $\alpha=1$ initial bursting is firstly observed at $\beta = 0.8$. Fractal order $\alpha$ has more influence in the ceasing of bursting, it is possible to generate initial bursting with a higher value of $\alpha$. The bursting inferior boundary is given by (blue dotted line) 
\begin{equation}
\label{eq14}
q(\alpha) = 1.08\alpha-0.275,
\end{equation}
and the superior (purple dotted line) by
\begin{equation}
\label{eq15}
g(\alpha) = \alpha+0.059.
\end{equation}
Another transition occurs from Region III to I, where the boundary is defined by
\begin{equation}
 r(\alpha) = 1.24\alpha+0.003.
\end{equation}

\begin{figure}[htb!]
    \centering
    \includegraphics[scale=0.35]{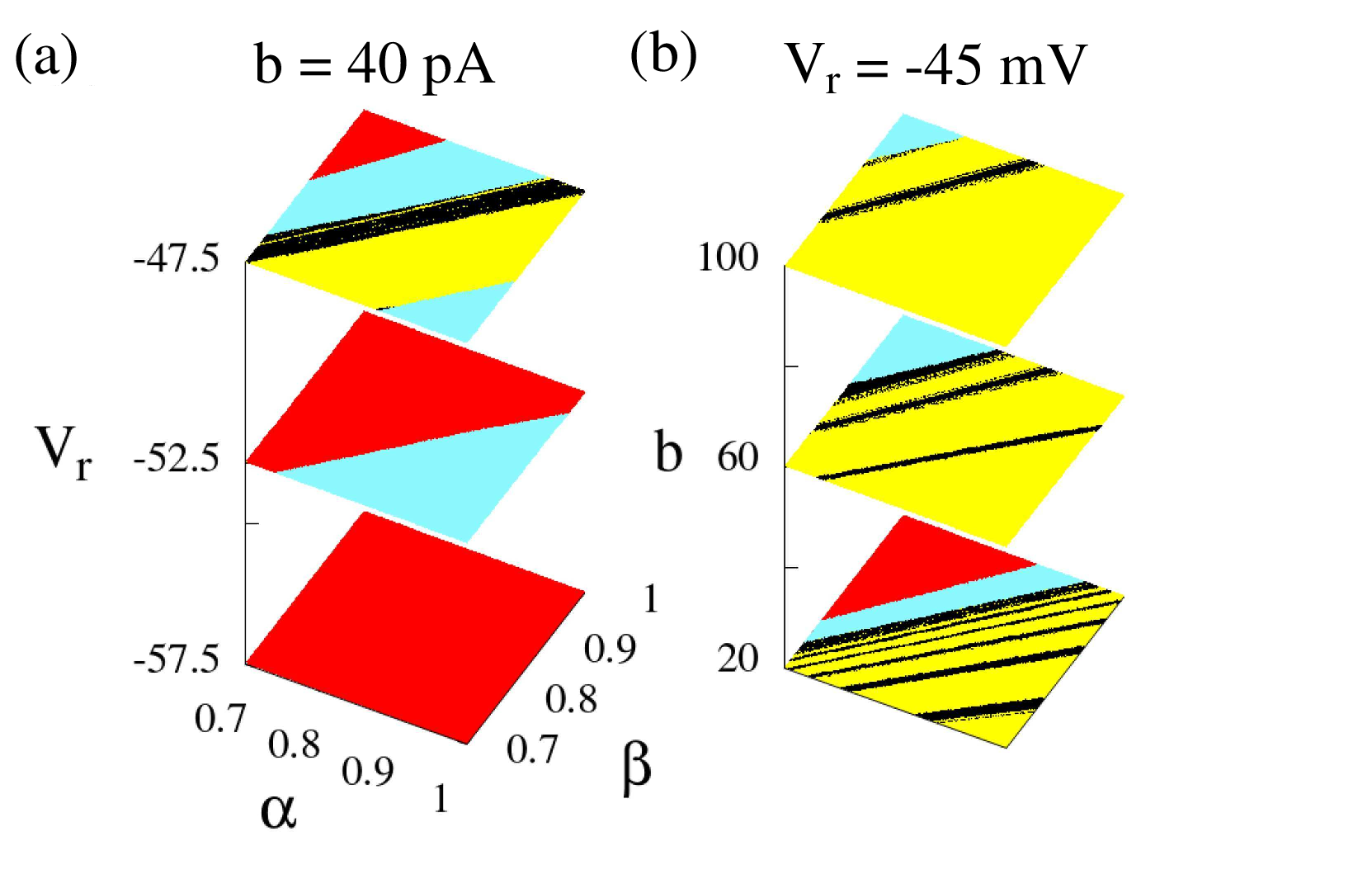}
    \caption{$\alpha \times \beta$ parameter space for different combinations of reset parameters. (a) We fix the reset parameter $b$ = 40 pA and vary $V_{\rm r}$ and (b) we fix $V_{\rm r} = -45$ mV and vary $b$.}
    \label{Fig8}
\end{figure}

The reset parameters influence how $\alpha$ and $\beta$ change the firing patterns, Figure \ref{Fig8}. The panel (a) shows $\alpha \times \beta$ parameter space for three different values of $V_{\rm r}$. Small increments of this reset parameter add more firing pattern regions. For $V_{\rm r}= -57.5$ only adaptive spiking is seen (red region). For $V_{\rm r} = -52.5$, two regions are seen, the red region  (adaptive) and the blue region (initial bursting). Considering $V_{\rm r} = -47.5$, four firing patterns appear, red region (adaptive spiking), blue region (initial bursting), yellow region (regular bursting) and black region (irregular bursting). The variation of the reset potential influences the initial, regular and irregular bursting appearance in ($\alpha,\beta$) space parameter. The panel (b) shows ($\alpha, \beta$) parameter spaces for three values of b for $V_{\rm r} = -45$ mV. $b$ increament has the opposite effect of $V_{\rm r}$ increase, higher values of $b$ reduces the amount of firing patterns seen in the parameter space. For $b=20$ pA, four different firing patters regions are shown, red region is the adaptive spiking, blue region initial bursting, yellow region regular bursting and black region irregular bursting. For $b=60$ and $b=100$ pA, only three firing patterns are generated. However, the irregular bursting pattern regions reduces for $b>20$ pA.

These results, combined with the ones presented in Figure \ref{Fig6}, reinforce that different firing patterns are not only produced by $V_{\rm r}$ and $b$ but also by the non-integer order of the extended operator. Both orders influence the model dynamics, the fractal order $\beta$ greatly affects the adaptability of the neuron. However, the potential membrane fractal order $\alpha$ has greater influences on the firing pattern transition of bursting to spiking.

\section{Discussion and Conclusion}
In this work, we proposed and studied the adaptive exponential integrate-and-fire (Adex) model with fractal extension. We noticed that the fractal order Adex is capable of reproducing neuronal activity as Adex standard model. However, differently from the standard one, the fractal order presents quantitative differences in the adaptability, firing frequency, and variability of the inter-spike intervals. When the same fractal order is considered in both variable evolution related to the membrane potential ($V$) and adaptation current ($w$), the reduction of the fractal order in the membrane potential derivative reduces the tonic spike firing pattern in the studied parameters space. The reduction of the fractal order works to make the spike firing difficult, i.e., when $\alpha$ is reduced the firing frequency becomes lower. Moreover, our results also suggest that the mean inter-spike intervals follow an exponential law as a $\alpha$ function. 

In addition, when the fractal order is considered independently in $V$ and $w$, we observed that the reduction of the fractal order in the adaptation current generated a larger effect than in the potential variable, increasing the adaptability in the firing patterns. In our simulations, the reduction in the fractal order in an independent way produces similar magnitude changes in the coefficient of variation and mean firing frequency. In particular, we highlight the transition from tonic spike to adaptive spiking due to the reduction of the fractal order. We also concluded that, depending on the combination of fractal order in the variables, the model can exhibit other patterns, however, as the main result, the tonic spike is reduced or absent as a consequence of order reduction.

Eqs. (15) and (17) are very similar, their only difference is the independent term 0.088 in Eq. (15). Their slope is the same, which shows that the transition of the patterns in both cases occurs at the same rate of variation of the fractal orders. Considering $\alpha=0.7$, the transition of initial to regular bursting in Figure 7 (a) occurs at $\beta =  0.809$. The independent term in Eq. (15) plays a crucial role in the value in which this transition happens. The other boundaries are not as similar as Eqs. (15) and (17), their slopes and independent terms vary. These boundary equations limit the values of ($\alpha, \beta$) in which regular and irregular bursting are observable. Therefore, it is possible to select a combination of ($\alpha,\beta$) to describe the desired firing pattern. In future works, we plan to study a way of generalization of the bursting boundaries for any reset parameters to expand the applicability of these equations.

When fractal derivatives are considered, their orders influence the firing pattern similarly to the reset parameters in the standard model. Two sets of parameters are necessary to achieve different firing patterns, the reset parameters ($V_{\rm r}, b$) and the fractal orders ($\alpha, \beta$). ($V_{\rm r}, b$) change the firing pattern parameter space of $\alpha \times \beta$ and vice-versa. However, the parameters act differently in each parameter space. ($\alpha, \beta$) increase the mean interspike interval and suppress the emergence of tonic spiking, and the reset parameter $b$ reduces the amount of firing pattern shown in $\alpha \times \beta$ parameter space. 

The AdEx model has the advantage of using few parameters to reproduce experimental data. In particular, it is possible to obtain six different types of firing patterns by varying just two parameters ($V_{\rm r}$ and $b$). The parameter $b$ is related to adaptation and can be varied to better represent the temporal evolution of ISIs of a neuron. On the other hand, the parameter $V_{\rm r}$ estimates the value of the membrane potential after an action potential. This parameter can be measured directly from voltage traces and cannot be arbitrarily changed.
It is observed that cells with the same value of $V_{\rm r}$ can exhibit different adaptation levels, firing patterns and frequencies$^{16}$. For this reason, it is necessary to include new parameters to accommodate this firing diversity. We show how the Adex model with fractal extension enables a better representation of the adaptability, variability, and frequency of firing patterns, in this way expanding the applicability of the Adex model.
In future works, we plan to use the fractal orders as a mean of fitting experimental data of single neurons to neuronal networks, expanding the model and its applicability for more general situations and also to better understand some brain disorders.
 
With respect to the order of fractal derivatives, our study also encountered some limitations. The meaningful solutions in terms of biological plausibility are found in the range $\alpha \in (0.7,1]$. Although our results are revealing regarding the potential effect of fractal order in single neurons, more investigation into the effect of fractal order on the dynamics of neuronal networks is necessary. In conclusion, our study provides clear evidence of the influence of the fractal order in the firing patterns and frequency of the Adex neuron model. Our results expand the applicability of the Adex model, supporting future investigations considering the fractal order neuron models.

\begin{acknowledgments}
The authors acknowledge the financial support from S{\~a}o Paulo Research Foundation (FAPESP, Brazil) (Grants N. 2018/03211-6, 2020/04624-2 and 2023/12863-5), support from the Brazilian Federal Agencies (CNPq) under Grant No. 304616/2021-4, and D.L.M.S. received partial financial support from Coordena{\c c}{\~a}o de Aperfei{\c c}oamento de Pessoal de N{\'i}vel Superior - Brazil (CAPES) grant number 88887.849164/2023-00. E.C.G. received partial financial support from Coordena{\c c}{\~a}o de Aperfei{\c c}oamento de Pessoal de N{\'i}vel Superior - Brasil (CAPES) - Finance Code 88881.846051/2023-01.
\end{acknowledgments}

\section*{Data Availability Statement}
The authors confirm that the data supporting the findings of this study are available within the article.

\section*{References}
\vspace{-0.5cm}
\nocite{*}
\bibliography{aipsamp}
\end{document}